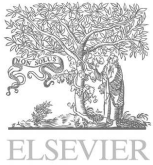

Journal logo

# AMICA, an astro-mapper for AMS

Alessandro Monfardini,[a*] Paolo Trampus,[b] Roberto Battiston,[c] Corrado Gargiulo[d]

[a]*INFN gruppo di Trento and ITC-IRST, via Sommarive 18, 38050 Povo (TN), Italy*

[b]*Center for Advanced Research in Space Optics, Area Science Park, Padriciano 99, 34012 Trieste, Italy*

[c]*INFN sez. Perugia and Università di Perugia, Dipartimento di Fisica, via Pascoli, 06123 Perugia, Italy*

[d]*INFN sez. Roma 1, P.le A. Moro 2, 00185 Roma, Italy*

**Abstract**

The Alpha Magnetic Spectrograph (AMS) is a composite particle detector to be accommodated on the International Space Station (ISS) in 2008. AMS is mainly devoted to galactic, charged Cosmic Rays studies, Antimatter and Dark Matter searches. Besides the main, classical physics goals, capabilities in the field of GeV and multi-GeV gamma astrophysics have been established and are under investigation by a number of groups. Due to the unsteadiness of the ISS platform, a star-mapper device is required in order to fully exploit the intrinsic arc-min angular resolution provided by the Silicon Tracker. A star-mapper is conceptually an imaging, optical instrument able to autonomously recognize a stellar field and to calculate its own orientation with respect to an inertial reference frame. AMICA (Astro Mapper for Instruments Check of Attitude) on AMS is responsible for providing real-time information that is going to be used off-line for compensating the large uncertainties in the ISS flight attitude and the structural degrees of freedom. In this paper, we describe in detail the AMICA sub-system, the accommodation/integration issues and the in-flight alignment procedure adopting identified galactic (pulsars) and extragalactic (AGNs) sources. © 2001 Elsevier Science. All rights reserved

*Keywords*: space instrumentation, gamma-ray astronomy, star-tracker, International Space Station.

*PACS*: 07.87.+v (Spaceborne instruments), 07.60.-j (Optical instruments), 95.55.Ka (X and gamma telescopes), 95.55.-n (Astronomical and space-research instrumentation).

---

[*] Corresponding author. Tel. +39 0461 314258  Fax +39 0461 314340. e-mail: monfardini@itc.it

## 1. Introduction and science case

AMS [1] is a complete particle detector to be flown for at least three years on-board the International Space Station (ISS). The launch is scheduled on the UF-4.1 ISS construction Space Shuttle flight. The AMS physics program [2] includes Antimatter search with unprecedented sensitivity, a quantitative analysis of the galactic Cosmic Rays, Dark Matter indirect searches and Gamma-Ray astrophysics. The Gamma Ray physics domain will be probed by AMS using two different and complementary approaches [3]:
- by measuring converted $e^+/e^-$ pairs in the Silicon Tracker (conversion mode);
- as single photons in the imaging Electromagnetic Calorimeter (single-photon mode).

The photons interacting directly in the Electromagnetic Calorimeter (single-photon mode) are traced back to the original direction with a poor angular resolution of the order of 1÷3 deg. On the other hand, taking advantage of the excellent Silicon Tracker intrinsic resolution, in conversion mode the single gamma photon is much better reconstructed. More precisely, the angular resolution improves with increasing $E_\gamma$. For $E_\gamma > 30$ GeV the angular uncertainty vs. energy curve reaches a plateau at the "astronomical significant" value of 1÷2 arcmin (0.3÷0.6 mrad). It is important to point out that this is the angular uncertainty associated to a single photon; the final source localization box size will depend on the brightness/spectrum combination and will be smaller. About half of the overall AMS gamma throughput for $E_\gamma > 1$ GeV (~1000 cm$^2 \times$sr) is related to the high angular resolution conversion mode. Please refer to [3,4] for a more detailed description of AMS as a gamma-ray telescope.

The aim of the AMICA star-mapper is to provide a precise ( < 20 arcsec), real-time 3D transformation of the AMS mechanical x-y-z frame to sky coordinates regardless of the large uncertainties introduced by the ISS attitude behaviour and structural elasticity. That transformation error, if uncorrected, would definitely dominate the intrinsic Silicon Tracker (TK) one even on a single photon basis and destroy the source identification capabilities. In other words, it is the combination star-mapper and Silicon Tracker that guarantees a real, sky-projected, angular resolution in the astronomical range of interest.

The AMS-γ exposure/resolution performances will allow clean studies of the known galactic (e.g. Pulsars) and extra-galactic (e.g. Blazars and AGNs in general) sources, as well as the potential discovery of new multi-GeV emitters. Particularly interesting are the Gamma Ray Bursts (GRBs), and among them the still mysterious short/hard population. Provided the persevering large electromagnetic spectrum coverage of GRBs phenomena, the source compactness and the explosive nature of the event, observation in the multi-GeV domain would end-up in a precise estimate of a potential photon speed energy dependence [5].

## 2. The AMICA star-mapper

The AMICA star-mapper (AST) is an evolution of an existing pointing/tracking system flown three times on the Space Shuttle as part of the UVSTAR telescope [6,7].

A star-mapper is conceptually a smart optical instrument pointed to the sky and with sufficient field-of-view and sensitivity to actually see a certain number of stars per exposure. The observed point sources are then optimally matched with an on-board astrometric/photometric catalogue. The final result is a determination of the instantaneous orientation with respect to an inertial reference frame. The real mapping system complexity strongly depends on the required performances (duty cycle, astrometric error, mapping frequency) and on the operational environment (day/night, background sources, thermal, radiation, electronic interface etc.). According to the system requirements for AMICA on AMS, the performances are:
   a) high day/night duty cycle;
   b) no moving parts;
   c) residual pointing error: few arcsec intrinsic [7]. A comparable, additional contribution is then introduced by the long supports finite stiffness and by the thermal deformations (see next paragraph);
   d) maximum attitude determination rate: 15 Hz;

e) dedicated interface with the AMS Universal Slow Control Module (USMC);

f) on-board ms clock periodically synchronized by the external GPS unit with ms precision;

g) total mass: 8 kg including cables and supports;

h) power budget: 18 W;

i) complicate geometrical and pointing constrains accomplished.

The high projected angular rate of the ISS (~240 arcsec/sec) forces the optical detector to work in fast shots mode. The light collecting area, on the other hand, is limited by geometrical/mass constrains to 28cm$^2$. An internal-gain 12-bit camera has been developed and built to reconcile accounts. The calculated visual magnitude detection limit is at least V=7.5 considering an exposure time of 40 ms. The core of the device is an L3 vision technology, auto-intensified, frame-transfer CCD chip (format: 512×512, pixel size: 16μm). The gain is achieved by means of an internal low-voltage multiplication register and can be set between 1 and 1000. The 12-bits effective digitisation covers the entire dynamic range between the brightest stars and the detection limit. Two such cameras (named ASTC1 and ASTC2) are controlled by a common, central DSP-based board (ASTEP) in charge of performing the image extraction, fitting the star positions to the on-board catalogue, calculating the AMS attitude. The serial line communication to AMS are also provided by ASTEP. In order to ensure a safe redundancy, the central board is connected to two USCM modules. The two cameras are connected to the processor board through 3.5m long cables. The electrical protocol is based on the spacewire bus. The commands and data exchange are based on proprietary protocol.

The ASTEP hardware has two image buffers, each one dedicated to the corresponding camera. The fast serial data coming from the cameras are transferred directly into the image memory. Once available, a program locates/extracts the stars from the image and calculates their centroid positions and relative intensities. All the values are temporarily stored in a table (observed stars table). The system now loads a portion of the on-board stellar catalogue and projects it onto the CCD plane (catalogue stars table) for comparison.

Two further tables containing the inter-angles are generated, and an hypothesis of correspondence between observed and catalogue stars is performed. The orientation algorithm accepts or rejects this hypothesis based on the inter-angles tables comparison. In case of a rejection another hypothesis of correspondence is made, and the procedure is repeated until the acceptance (matching) is reached. The algorithm then minimizes the distances between the acquired stars and the matched ones and determines the orientation of three reference axes (the camera pointing axis and the two CCD sides). A matrix multiplication allows at the end to reconstruct the AMS tracker pointing direction at the time when the image was acquired. The initial pointing (e.g. at power-on) is determined through a dedicated all-sky searching algorithm and/or by calculating the approximate ISS orbit starting from the orbital parameters given from ground.

Another electronic board (ASTEI) is responsible for distributing power to the subsystems. Both ASTEP and ASTEI are allocated in a VME-like crate. At present, all the function are integrated in a single (ASTEP) board.

The AMICA optical system consists in a special, fast lens (75mm, f/1.25). The L3 CCD detector will sample the inner, higher quality 6.25×6.25 deg field-of-view with a plane scale of 44 arcsec/pixel and with a time rate between 10Hz and 30Hz. The back focal length is small (2÷3 mm) allowing an extremely compact design. The lens has been tested, in vacuum environment, up to a temperature of 60°C. An ad-hoc composite optical filter (GG475+KG2) is mounted in front of the input lens in order to maximize the S/N ratio against scattered light and to limit the input power.

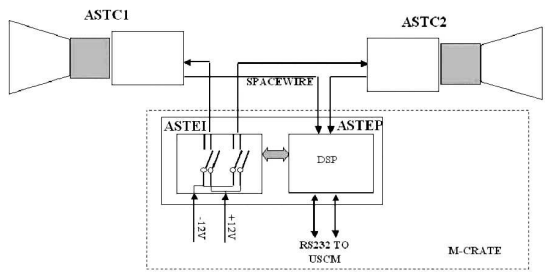

Figure 1. AMICA electrical connections scheme.

A number of tests have been carried out on a specific optical bench. The lens was able to achieve a sub-pixel Point Spread Function (PSF). However, in order to actually achieve sub-pixel precision via barycentric technique, we are going to degrade it in a controlled way. The optimal defocusing offset to be applied to the CCD was determined and is about 150μm (FWHM ~ 3 pixels). The distortions across the focal plane, negligible according to the specifications, have been experimentally constrained to less than 0.2 pixels. The upper limit is mainly determined by the step-to-step jitter of the adopted motors. As a final test, the lens intrinsic baffling efficiency has been measured and is reported in figure 3.

An external, double-cone Aluminium baffle has been designed to suppress background stray light to an acceptable level. The internal geometry, fixed according to the results of ray-tracing Montecarlo simulations, is schematically shown in figure 2. The "inner acceptance" angle (14 deg) is the maximum inclination for a ray to enter directly into the lens, while the "ultimate acceptance" (78 deg) angle is the minimum inclination to illuminate at least one inner edge. The Sun avoidance angle depends on the characteristics of the internal coating, and lie between these extreme values. The Moon avoidance angle, on the other hand, is expected to be smaller and close to the inner acceptance.

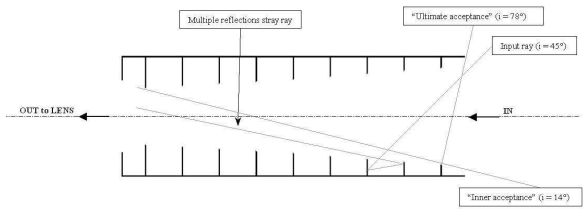

Figure 2. Baffle internal geometry.

A prototype baffle, with the final internal geometry, has been shaped and tested. A provisional, regular black coating has been internally grown by anodisation. The coating visible Total Hemispheric Reflectance (THR) has been estimated to be at least 5%, while a direct measurement of the scattered light yielded up a $10^{-5}$ suppression factor for a 40 deg incidence angle. By combining this measured figure with a conservative estimate of the internal baffling contribution, we expect an overall rejection figure of about $10^{-7} \div 10^{-8}$ and uniform distribution all over the CCD area. Other, more efficient types of coatings are under consideration. The simulation results are plotted in figure 3 assuming a THR of about 1%. According to the minimum acceptable detection limit ($m_V = 6$), we estimate a Sun avoidance angle of 45 deg.

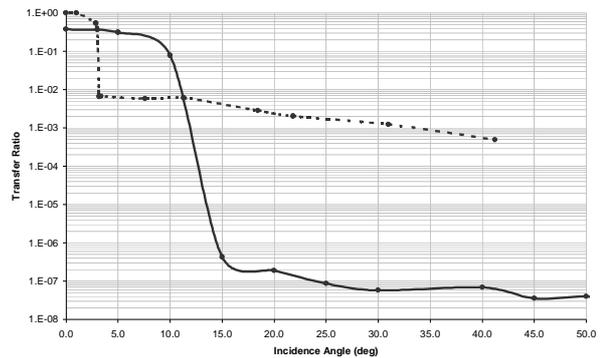

Fig. 3. Simulated external baffle efficiency (solid line). The transfer ratio is defined as the ratio between the power transferred to the lens and that entering the baffle. The dotted line represents the measured lens intrinsic baffling efficiency (internal baffle efficiency).

## 3. ISS environment and accommodation issues

The AMICA subsystem has been included in the AMS design when the latter was already in a relatively advanced stage. Furthermore, AMS is an extremely complex detector, and the interfacing study required a number of interactions and trade-offs.

At least two, well separated sky footprints are required to account for the presence of the Sun and the Moon. The minimum separation angle is determined by the Sun avoidance angle, that is itself a direct consequence of the resources available for the baffling system. The pointing directions of the cameras should be as far as possible, free of obstructions from ISS fixed or moving parts and sufficiently far from reflecting surfaces.

It has been pointed out that to ensure a reliable mapping the pointing directions have to remain fixed

(within the specified errors ~ arcsec) with respect to each other, and each one with respect to the Silicon Tracker axes. Due to the severe thermal gradients expected on the exposed surfaces, the only way to achieve it is to attach the ASTCs directly to the temperature-controlled Silicon Tracker main frame. This last issue limits the pointing directions choice and sets an upper limit on the achievable duty cycle.

Finite elements mechanical simulations has been carried out to establish the pointing stability in AMS-ISS typical thermal/mechanical conditions. The adopted solution, requiring sophisticated carbon fibre supports, is illustrated in figure 4. The achieved stability is of the order of arc-sec, basically negligible in the overall error budget. The two cameras are identified as "starboard" (ISS right wing) and "inport" (ISS left wing) according to the ISS technical conventions. The optical axes have both a 50 deg elevation with respect to the AMS own xy plane (perpendicular to the AMS pointing direction).

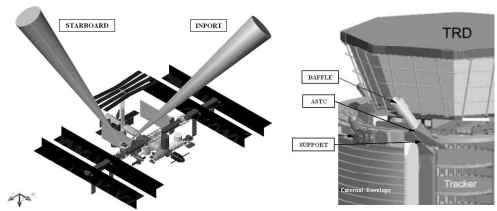

Figure 4. Left: ASTCs field-of-view and the ISS structure. Right: particular of an ASTC camera.

The starboard camera, facing the outward side of the ISS, can't avoid a ~10% average time obscuration determined by the ISS rotating solar panels. A significant seasonal jitter is expected.

As far as the inport device is concerned, the default attitude of the ISS (about 10 deg "nose down", 5÷10 deg "nose left") and the AMS tilting (12 deg "inward") contribute to lower the effective elevation over the Earth airglow. Assuming a typical airglow height of 122 km, a conservative estimate of the real elevation is 34 deg. Furthermore, the inport camera will periodically suffer from solar panels reflected sunlight.

The ASTCs viewing angles have been fixed as a result of a trade-off study. Realistic views from the two AMICA cameras are shown in figure 5 for two particular solar panels configurations.

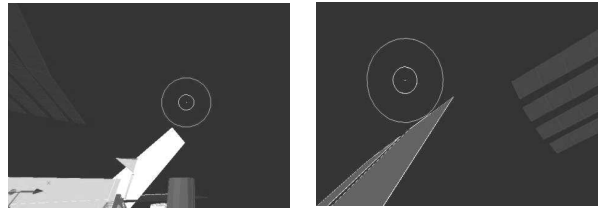

Figure 5. Left: ASTCs field-of-view and the ISS structure. Right: particular of an ASTC camera.

A detailed visibility study has been performed assuming 45 deg and 20 deg respectively for the Sun and the Moon avoidance angles. One year ISS orbital time has been simulated assuming the nominal attitude. A conservative statistical correction is added to take into account the obscuration/reflection periods. The resulting estimation of the overall system "environmental" duty cycle is 0.94÷0.97 (best÷worst season). These values are strongly dependent upon the Sun avoidance angle and the first avoidance angle. The correction introduced by the finite failure rate of the matching algorithm itself is of the order of 0.1% and is thus negligible.

We extensively investigated the radiation environment, paying particular attention to the CCD silicon chip. A dedicated test run has been performed at the LNL SIRAD irradiation facility to qualify the device for space applications [8]. To briefly summarise the results, we note that the large shielding provided by the AMS structure limits the Total Ionizing Dose (TID) to the quite safe level of about 50 rad(Si)/year over the CCD. On the other hand, the destructive Single Events Effects (SEE) rate has been experimentally estimated to be less that $6.6 \times 10^{-4}$ $y^{-1}$ and thus negligible.

The steady magnetic field generated by the AMS superconducting magnet has been taken into account. Even if a field strength of 2100 gauss has been estimated at the location of the two AMICA cameras, no significant effects are expected on the electronics components.

## 4. In-flight alignment procedure

The relative orientation of the Silicon Tracker and ASTCs cameras will be determined in laboratory with optical techniques. The rotations have in any case to be cross-checked in orbit to verify that no fixed offsets have been mechanically induced during the launch/deployment phases. Possible periodic or reversible displacements are negligible as explained in the previous paragraph.

The Silicon Tracker must thus be able to get few astrometric references in sky by intercepting known and optically identified gamma sources. At the same time, by comparing the simultaneous ASTCs determinations, the rotation matrixes will be determined on a single photon basis. The most powerful galactic/extragalactic GeV emitters are the obvious candidates. Among them, we conservatively restricted our analysis to three pulsars (Vela, Crab, Geminga) and to the 3C279 Active Galactic Nucleus (AGN) located at moderate redshift ($z \approx 0.5$). A Montecarlo procedure has been developed to properly take into account the combined energy/angular resolution of the Silicon Tracker and the spectral behaviours of the sources. Starting from the EGRET [9] spectral indexes, we prudentially adopted the polar cap model [10] for the pulsars and introduced an EBL (Extragalactic Background Light) attenuation term for 3C279 [11]. The spectra have been calculated up to 100GeV, and the weighted (with the angular resolution) photon distribution generated.

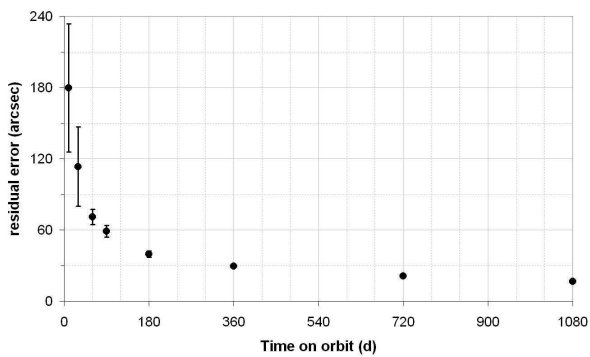

Figure 6. Expected residual alignment error (68% containment radius) versus time in an averaged orbit.

The resulting alignment residual error resulting from the Montecarlo is reported in figure 6 as a function of time for the mean orbit. Of course, strong seasonal variations are expected in real orbit operation. The exact figure could be determined only after the exact launch date and ISS attitude strategy will become available.

On average, we expect the residual error to become smaller than the tracker best angular resolution (1 arc-min) after 3 months. In a 3 years mission, on the other hand, the residual alignment error will be as small as 15 arc-sec, negligible on a single photon basis. It is important to point out here that every single photon coordinates could then be corrected a posteriori with the final residual error. In other words, the alignment process is not going to limit in any way the astrophysical capabilities of AMS, even in the very first months.

## 5. Conclusions

We described the design, expected system performances and laboratory test results of the AMICA star-mapper. AMICA will significantly contribute in extending the AMS capabilities into the field of high-energy gamma astrophysics as explained in the first paragraph.

We then gave a moderately detailed description of the star-mapper itself, touching all the involved subsystems: cameras, DSP electronic board, matching software, optics and external baffle.

Lastly, we reported the accommodation limitations due to a number of mechanical and resources constrains. Despite the forced non-optimal pointing directions of the two cameras and the severe limitation imposed on the baffling system, the simulated duty cycle over one year of observations exceeds 94%, while attitude determination capabilities are well within the required 20 arc-sec. The suitability of the on-flight astrometric calibration procedure has been demonstrated via Montecarlo simulations applying prudent high-energy emission cut-offs on both Pulsars and AGNs known objects classes.

## Acknowledgments

We acknowledge the important contribution of Dr. Agnieszka Jacholkowska for the calculation of the absolute gamma-ray fluxes expected in AMS.